# Genomics and Biological Big Data: Facing Current and Future Challenges around Data and Software Sharing and Reproducibility

Sandra Gesing
Center for Research Computing
University of Notre Dame
Notre Dame, USA
sandra.gesing@nd.edu

Thomas Richard Connor
Cardiff University School of Biosciences
Cardiff, UK
ConnorTR@cardiff.ac.uk

Ian Taylor
University of Notre Dame, USA
And Cardiff University, UK
itaylor1@nd.edu

*Abstract*—Novel technologies in genomics allow creating data in exascale dimension with relatively minor effort of human and laboratory and thus monetary resources compared to capabilities only a decade ago. While the availability of this data salvage to find answers for research questions, which would not have been feasible before, maybe even not feasible to ask before, the amount of data creates new challenges, which obviously need new software and data management systems. Such new solutions have to consider integrative approaches, which are not only considering the effectiveness and efficiency of data processing but improve reusability, reproducibility and usability especially tailored to the target user communities of genomic big data. In our opinion, current solutions tackle part of the challenges and have each their strengths but lack to provide a complete solution. We present in this paper the key challenges and the characteristics cutting-edge developments should possess for fulfilling the needs of the user communities to allow for seamless sharing and data analysis on a large scale.

## I. INTRODUCTION

The last 10 years has seen an explosion in the quantity, variety and complexity of data that is routinely generated by researchers examining questions in the life sciences. This rise has been driven by the development of new high-throughput technologies, which are perhaps best exemplified by those that currently dominate genomics. The progenitor of the sequencing instruments, the Genome Analyser [1], was launched by Solexa (now Illumina) in 2006 and now powers the majority of the genomics revolution. Immediately prior to the release of this technology, the human genome had taken 15 years and over $2 billion to complete [2]. It is now possible, using a direct descendent of the original Solexa instrument, to complete the same task in < 3 days for less than $1000 [3]. The capacity to rapidly and routinely unravel the genetic code for entire organisms has opened new frontiers in biology, changing the sort of research questions that can be asked, and the scales at which they can be examined. However, while mechanisms exist for sharing the data generated along with analysis packages themselves, these systems - founded upon ideas of biological gateways, tools and databases founded and templated on pioneers that emerged in the 1990's [4], 80's [5] and 70's [6] respectively - are now contributing to a data and software searchability, reusability and reproducibility crisis that leads to the paradoxical situation where biological data and software are often shared, but rarely reused nor results can be reproduced. Examples are presented in studies in medicine and pharmacology with the outcome that only 11% [7] or 6% [8] of the analysed research was reproducible. In general, there is a large gap knowledge and experience in the research domain and in data services between data scientists or computer scientists and domain researchers. The vision was to bridge this gap via bioinformaticians as knowledgeable intermediaries between the two worlds. While bioinformaticians have been proven to significantly enhance the uptake of IT methods in biology, the disadvantage is that domain researchers are less aware of IT capabilities and concepts that they can apply to their specific research question and which make their IT-related work more effective and efficient. This situation poses a major challenge to biological big data analysis, and we believe that the solution to the nascent data and software reuse crisis is to be found through research around how we better store and share biological data and software and increase the usability of services to access and process the data.

In this paper we will specifically outline some of the key challenges facing the life sciences around data and software sharing, that collectively imperil the enormous potential benefit that could be reaped from the generation, sharing and reuse of software and data generated by thousands of life sciences researchers around the world. We then move on to consider the future areas for research that will be required in order to deal with these growing issues in the life sciences.

## II. CURRENT APPROACHES

Following on from the Human Genome Project, there are now a wide range of databases and software tools available for biologists who wish to examine experimental data. These databases are designed to store large quantities of structured data of a particular type, and include significant international resources such as the European Nucelotide Archive [9] and Uniprot [10]. However, many of these systems are evolutions of systems that were designed to store the quantities of data that were being generated before the advent of next generation sequencing. Many of these databases provide APIs that enable an experienced user to interrogate them via software [11] - but for the majority of users, the interfaces, built around storing and accessing single genomes, provide a significant barrier to their use. Compounding this design issue, most research datasets now comprise multiple, related data types, and as a result these focused databases do not provide an efficient mechanism for users to rapidly search, find and access all of the data for a project interest. They also provide few tools to integrate across distributed databases in order to reconstitute and combine published research datasets. Introductions of ideas such as the "Bioproject" at NCBI [12] seek to partially address this problem, however, these solutions remain limited to data across a small number of databases, hosted by NCBI. While relational databases have been often the choice for providing data (e.g., MariaDB [13]), novel concepts with NoSQL [14] approaches overcome limitations of those databases with a more suitable method for web-based data management, cloud applicability and performance optimization for big data. However, the usability is often very limited and not self-explanatory for the users.

The crossover between intuitive user interfaces with access to sophisticated tools extending websites with access to large databases are missing or are heavily overused (e.g., RAST [15]) or are too inflexible for seamless adaption to further use cases by end users.

III. USABILITY AND REPRODUCIBILITY VIA WORKFLOWS AND SCIENCE GATEWAYS

Since the user community in bioinformatics and genomics and other user communities processing data on a large scale are not mainly consisting of data scientists but of domain scientists, a wide range of mature science gateways and workflow systems have been developed in the last decade to increase the usability of tools and data via intuitive user interfaces. Additionally, the reproducibility can be assured via pre-defined workflows suitable for analysis steps in a defined order based on control and data dependencies for data on a large scale. Such science gateways and workflow systems include Galaxy [16], Pegasus [17], KNIME [18], Taverna [19], Kepler [20], Swift [21] and WS-PGRADE [22], to name a few widely used in the life sciences domain, and their key goals include reproducibility and reusability. They have large user communities with specific preferences for diverse technologies, capabilities and IT skills. Thus, each of them has advantages and disadvantages regarding the target user base, the way data is processed, diversity of data sources and presentation of results. They can be distinguished via their diverse workflow concepts, the diverse supported workflow constructs (e.g., DAG-based, additionally loop constructs or parameter sweeps) and can be utilised through web-based services (i.e. an API), as a graphical workbench necessitating installation on the users' side, or a combination of both, by providing a web-based downloadable installation mechanism (e.g. Web Start for WS-PGRADE). To date, there are no graphical full-featured web-based applications that provide a workflow environment to avoid further programming or installation for the user. Science gateway and workflow systems providers therefore have realised additional to the need for reproducibility and reusability also the need for usability to improve the users' experience and target their specific needs evident in the growing number of graphical user interfaces and social science platforms. However, internal as well external dependencies on operating systems, tools in diverse versions and local or distributed data hamper this goal. A study on the social marketplace MyExperiment [23] for sharing Taverna workflows, for example, illustrates that 80% of the workflows are not reproducible and not reusable out of the box [23]. Workflow interoperability between diverse workflow systems even increases the complexity and approaches such as the SHIWA course- and fine-grained workflow interoperability used in ER-flow [24] or interoperability between neuGRID and the Virtual Imaging Platform [25], can handle some conversions of workflows automatically but necessitate still manual steps to achieve a complete reusability and reproducibility.

IV. TOWARDS THE CLOUD

While science gateways, such as Galaxy, are well established as a mechanism for users to interact with their data, the advent of large cloud computing infrastructures for research, such as iPlant [26], NECTAR [27] and CLIMB [28], has resulted in the development of a new class of scientific gateway - the personal research gateway, which provide more customised interfaces, such as a VPS (Virtual Private Server), to its users. Within iPlant, for example, users are presented with the option of instantiating a VM from a predefined library, or interacting with their data through a "discovery environment". Within NECTAR and CLIMB, users can also access a self-service system to launch a custom VM, but are also presented with the option of creating a personal research gateway [29], which launches a dedicated VM running a set of predefined analysis services that the user can interact with via a website hosted on the VM. These solutions, running on top of cloud infrastructures, provide a steppingstone between classical single-site databases and associated analysis tools, and local research infrastructure and analysis. However, while they exist within a shared environment, they remain highly silo-ised, and data/software sharing within these infrastructures remains non-trivial for many users. Furthermore, while approaches such as the discovery environment or a personal research gateway may provide key tools for users, these are often service/cloud specific, limiting their spread and perpetuating a situation where resources exist for users to access, but there is little connectivity between the available resources to make data and software easily portable.

V. FUTURE CHALLENGES

The key limitations of current approaches are to be found in the fact that they are often disconnected from one another, and founded upon storing data of a very particular type, with limited connectivity to related data of different types. The cloud has considerable potential for empowering researchers to effectively share and reuse software and data. However, there are a number of research challenges that need to be met, before it will be possible to fully reap the rewards that are promised, but yet to be delivered, by cloud computing in the life sciences.

The first of these challenges is the challenge to integrate multiple disparate community clouds across the world into a single research infrastructure, where the user effectively does not have to worry about selecting the most appropriate cloud service for them. The advent of container environments, such as Docker [30] may play an important role here. Service selection should be transparent to the user - and they should be able to simply use the tools they need, share and access the data that they want, in as simple, rapid and limitation free way as possible. Users may want to share data not before they published or patented it. Thus, privacy and security needs have to be met in such an infrastructure but it should not hamper the sharing capabilities if they are desired. Additionally, a major drawback of cloud computing with an underlying business model is the costs for research groups. Users should have optimised exploitation of the resources suitable exactly to their use cases to minimise the costs.

The second of the key challenges is how we better integrate the cloud infrastructures with the principal data sources that are generated by researchers. Even though there have been developments for an integration of lab technologies, the state of the art are still disconnects between data sources and the efficient processing of data on distributed resources or a single research infrastructure as suggested above.

The third key challenge is how we combat the problem of software searchability, reproducibility and reusability. The diverse dependencies on hardware, operating systems, software versions and local or distributed data form a hurdle. While a single research infrastructure solves some of the problems, there is the need for concepts, which combine migration approaches with containerization approaches for execution environments capable of delivering a flexible solution easy to use by user communities.

Finally, there remains a key skills gap within life science researchers. Cloud computing provides the potential to train researchers in specific software toolsets, which, by hosting them in the cloud, will remain the same wherever a user accesses them from. For some time it has been suggested that researchers should simply upskill themselves, however, this is unrealistic, and has yet to yield any real results. Wetlab researchers are unlikely to want to invest the time in developing informatics skills that appear to be of relatively infrequent use to them, and to which they have little aptitude. Thus, the final key challenge is around how software and data are presented to users.